# ME-WARD: A multimodal ergonomic analysis tool for musculoskeletal risk assessment from inertial and video data in working places

Javier González-Alonso [1,*], Paula Martín-Tapia, David González-Ortega [2], Míriam Antón-Rodríguez [3], Francisco Javier Díaz-Pernas [4], Mario Martínez-Zarzuela [5]

*Department of Signal Theory, Communications and Telematics Engineering, Telecommunications Engineering School, University of Valladolid 47011 Valladolid, Spain*

A B S T R A C T

This study presents ME-WARD (*Multimodal Ergonomic Workplace Assessment and Risk from Data*), a novel system for ergonomic assessment and musculoskeletal risk evaluation that implements the Rapid Upper Limb Assessment (RULA) method. ME-WARD is designed to process joint angle data from motion capture systems, including inertial measurement unit (IMU)-based setups, and deep learning human body pose tracking models. The tool's flexibility enables ergonomic risk assessment using any system capable of reliably measuring joint angles, extending the applicability of RULA beyond proprietary setups. To validate its performance, the tool was tested in an industrial setting during the assembly of conveyor belts, which involved high-risk tasks such as inserting rods and pushing conveyor belt components. The experiments leveraged gold standard IMU systems alongside a state-of-the-art monocular 3D pose estimation system. The results confirmed that ME-WARD produces reliable RULA scores that closely align with IMU-derived metrics for flexion-dominated movements and comparable performance with the monocular system, despite limitations in tracking lateral and rotational motions. This work highlights the potential of integrating multiple motion capture technologies into a unified and accessible ergonomic assessment pipeline. By supporting diverse input sources, including low-cost video-based systems, the proposed multimodal approach offers a scalable, cost-effective solution for ergonomic assessments, paving the way for broader adoption in resource-constrained industrial environments.

## 1. Introduction

Work-related musculoskeletal disorders (WMSDs) are a major occupational health issue globally, particularly within the European Union. These disorders arise from ergonomic stressors such as repetitive motions, awkward postures, and biomechanical strain, which significantly impair worker productivity and well-being. Recent studies indicate that approximately 60 % of EU workers suffer from musculoskeletal disorders, with back pain (43 %) and muscular pain in the shoulders, neck, and upper limbs (41 %) being the most prevalent types (Eurofound, 2017). WMSDs affect employees across all sectors and professions, highlighting their prevalence in industrial settings (EU-OSHA, 2019). The economic burden is equally substantial, with WMSDs accounting for billions of euros in lost productivity and healthcare costs each year (Health and Safety Executive, 2021). Addressing these challenges requires innovative ergonomic assessment approaches aimed at reducing injury risks and improving workplace safety.

Traditional ergonomic assessments often rely on expert observations and the subjective application of assessment tools (Maldonado et al., 2015), such as the Rapid Upper Limb Assessment (RULA) method (McAtamney & Corlett, 1993). While effective, these methods are typically reactive—applied after injuries occur—and can lack precision, particularly in dynamic and complex work environments (Huang et al., 2020). This underscores the need for objective and reproducible tools capable of providing real-time ergonomic insights, especially in high-risk industrial workplaces.

* Corresponding author.
  *E-mail address:* javier.galonso@uva.es (J. González-Alonso).
[1] 0000-0003-3309-0578.
[2] 0000-0001-5271-1822.
[3] 0000-0002-3328-5183.
[4] 0000-0002-5625-9607.
[5] 0000-0002-6866-3316.

Motion capture technologies have emerged as a promising solution to address these limitations, while evaluating the effectiveness of ergonomic interventions over time (Menolotto et al., 2020; Vijayakumar and Choi, 2022; Ranavolo et al., 2018; Salisu et al., 2023). On one hand, Inertial Measurement Unit (IMU)-based systems, widely recognized for their precision, enable detailed tracking of joint angles and body movements without requiring an extensive setup (Sers et al., 2020; Carnevale et al., 2019; Poitras et al., 2019). These systems are particularly suited to industrial applications, where space constraints and flexibility are critical (Colim et al., 2021). Numerous studies have demonstrated that custom IMU-based systems—designed with affordability and flexibility in mind—can effectively meet the requirements for precise human motion tracking (Caputo et al., 2019; González-Alonso et al., 2021; Greco et al., 2020; López-Nava & Muñoz-Meléndez, 2016). Prior research (González-Alonso et al., 2024) has highlighted the feasibility of introducing motion data derived from custom or commercial IMUs into a computational framework for ergonomic assessment, enabling the automated calculation of reliable RULA scores in an automotive workstation. However, while IMU-based systems offer high precision (Colim et al., 2021; Huang et al., 2020), their performance can be affected by communication constraints, bandwidth limitations in real-time applications, and sensor drift over extended periods, reducing their practicality for long-duration assessments in real-world scenarios. Recent studies have addressed these challenges in communication constraints to optimize data transmission and improve system efficiency in constrained environments (Chiasson et al., 2020; Hou et al., 2016), while extensive calibration methods have been proposed to mitigate sensor drift (Maruyama et al., 2020).

On the other hand, multi-camera motion capture systems, such as Vicon (Oxford Metrics), are considered the gold standard in clinical and research settings for movement tracking due to their accuracy. However, their elevated cost, reliance on multiple cameras, need for positional markers, and spatial constraints limit their use in dynamic industrial environments (Panariello et al., 2022; Robert-Lachaine et al., 2017). Notably, recent advancements in computer vision, such as monocular 3D pose estimation systems—for instance, NVIDIA Maxine AR Body-Track (NVIDIA, 2021)—offer a cost-effective, markerless alternative, provided that workstation layouts are optimized to minimize occlusions and maximize tracking effectiveness (Ranavolo et al., 2018). Pretrained deep learning models for pose estimation in ergonomic assessment may face domain adaptation challenges similar to those in finger motion analytics or student and teacher models (Bigalke et al., 2023; Liu et al., 2021; Raychaudhuri et al., 2023). Additionally, deep learning-based object detection, known for its high precision in industrial small-object recognition (Li et al., 2024; Standley et al., 2017), could help automate some steps of ergonomic assessment, such as load quantification.

Although motion capture technologies are promising for ergonomic assessment (Menolotto et al., 2020; Vijayakumar and Choi, 2022), much of the existing research has been conducted within controlled laboratory environments, which do not fully capture the complexities of real industrial settings (Abobakr et al., 2019; Caputo et al., 2019). Studies, such as those by Abobakr et al. (2019) and Manghisi et al. (2017), have demonstrated the feasibility of RGB-D and depth camera-based systems for posture evaluation. However, their application in dynamic, resource-constrained industrial environments remains limited due to challenges related to occlusions, cost, and setup complexity.

In contrast to previous studies that primarily focus on a single type of sensor technology (Greco et al., 2020; Carnevale et al., 2019), the proposed ME-WARD tool offers several advantages over traditional and novel ergonomic assessment solutions in the literature, including its ability to seamlessly incorporate IMU or video-based motion capture systems in the ergonomic assessment processes of a factory. This multimodal capability provides greater flexibility in data acquisition, making the tool adaptable to different environments and budgets. A key advantage of the ME-WARD tool is its adaptability to real industrial environments, where workstations and operational conditions often vary significantly. Prior research, such as that by Panariello et al. (2022), has explored digital human models for ergonomic assessments, but these studies have been primarily restricted to static or simulated work conditions. Conversely, our approach leverages real-time data from active industrial processes, enabling dynamic monitoring and immediate intervention to mitigate ergonomic risks. The tool's flexibility enables ergonomic assessments across various data acquisition setups, enhancing the objective application of RULA. This research validates ME-WARD in industrial settings using a commercial IMU system and a monocular setting using a conventional camera, thus highlighting the current advantages and limitations of a state-of-the-art monocular body pose tracker for ergonomic assessment.

Ultimately, the ME-WARD tool provides an adaptable approach suitable for diverse ergonomic assessment contexts by enabling the integration of multiple motion capture technologies into a single, adaptable framework. Unlike existing solutions that are either cost-prohibitive or unsuitable for dynamic work environments, ME-WARD provides an affordable and flexible approach, supporting both high-end and low-cost setups. By enabling objective, real-time ergonomic monitoring, it facilitates proactive interventions, helping to reduce the risk of musculoskeletal disorders in a scalable and efficient manner.

## 2. Methods

### 2.1. Musculoskeletal risk assessment tool

The aim of this study is to introduce a digitalized approach to ergonomics risk assessment by utilizing different data acquisition setups to assess the risk of musculoskeletal injuries. The Rapid Upper Limb Assessment (RULA) method was selected to test the implementation due to its established efficacy in assessing ergonomic risk factors related to upper body postures, including those of the neck, back, and arms (Vignais et al., 2017). This method is particularly relevant for tasks involving repetitive or sustained movements, such as those commonly encountered in conveyor belt assembly work. By systematically scoring postures based on joint angles and body positioning, RULA serves as a reliable tool for identifying potential musculoskeletal risks, making it a cornerstone in preventive ergonomics research and workplace design improvements (McAtamney & Corlett, 1993).

In this study, we implemented a RULA computational method compatible with different motion acquisition technologies and evaluated its applicability by comparing it against the output generated by the commercial MVN MotionCloud RULA and Adjustable Ergonomics reports. First, we demonstrated that our customized approach produces comparable results using data obtained directly from Movella Awinda sensors. Subsequently, we applied the RULA computation to data collected from video-based systems. The input architecture of the developed tool (Fig. 1) builds upon and expands the pipeline established in a previous study (González-Alonso et al., 2024). That study conducted a comparative analysis between the RULA scores obtained from Movella's proprietary MVN Pro software and those generated using OpenSense (Al Borno et al., 2022), an OpenSim-based tool (Delp et al., 2007) employing a modified Rajagopal model (Rajagopal et al., 2016). By adding video-based input capabilities, the system not only strengthens the robustness of this digitalized ergonomic analysis approach but also enhances its versatility.

The proposed method consists of multiple stages, including data acquisition, preprocessing, ergonomic scoring, and result visualization. To improve overall performance, data filtering techniques, such as sensor fusion filters for IMU data and temporal smoothing for camera-based data, are applied to reduce noise and improve tracking stability.

In the approach proposed in this study, the RULA score can be computed directly from joint angles, eliminating the need to compute body parts positions. This method adheres to the traditional RULA framework by evaluating different body segments, including the arm, forearm, wrist, neck, and trunk, through a scoring system based on angular ranges and positions. The system also calculates the cumulative percentage of time that the subject remains within a given postural score range. Additionally, force/load scores, supported handling of heavy parts, and repeatability score are included as checkboxes in the application interface.

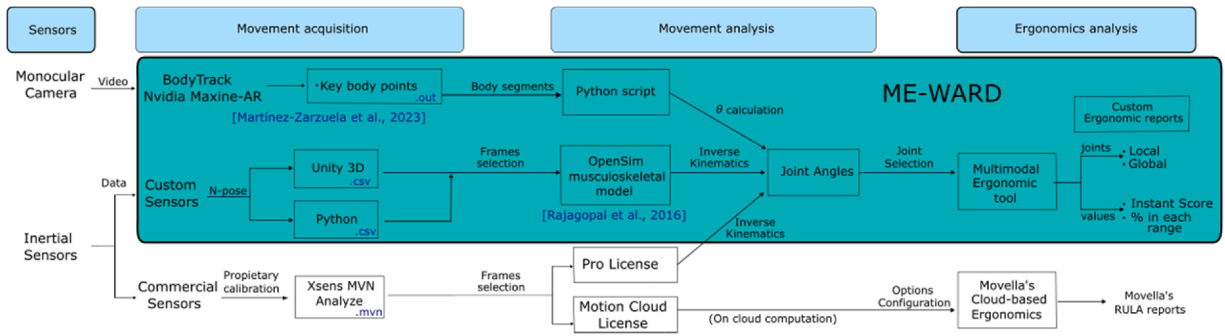

**Fig. 1.** Input scheme of the ME-WARD RULA pipeline.

The ME-WARD RULA analysis tool leverages reference tables (Tables A, B, and C) and predefined dictionaries that establish specific angular ranges for each joint along with their corresponding scores. These tables are derived from the original RULA method established by McAtamney and Corlett (1993) and provide standardized scoring criteria based on joint angles and postural data. Each table assigns a value based on angular position, ranges of motion, and body segment combinations, enabling the calculation of a total score that reflects the severity of ergonomic risk and facilitates the identification of priority areas for workplace interventions.

- **Table A** (see Table 1): Evaluates the upper body segments—upper arm, forearm, and wrist—alongside wrist twists, integrating adjustments for muscle use and load handling to compute the arm and wrist score.
- **Table B** (see Table 2): Focuses on the neck, trunk, and legs, incorporating similar adjustments for effort and load, and generates a corresponding posture score for these regions.
- **Table C** (see Table 3): Resolves the interaction between Table A and Table B scores, providing the final RULA score by combining the wrist/arm and neck/trunk/leg scores.

Each joint is assessed independently using a get-score function, which compares measured angles against the predefined limits in the scoring dictionaries. These scores are then integrated using the Table A and Table B functions, which account for interactions between body segments according to RULA's framework. Finally, the overall scores for the upper limbs and trunk/neck are combined via the RULA score function, using Table C to compute the total ergonomic risk score.

In the ME-WARD tool, these tables have been digitally implemented to automate the assessment process, ensuring compliance with the established RULA framework. The system is designed to accommodate additional ergonomic assessment methods by integrating corresponding tables tailored to specific evaluation criteria, offering flexibility for diverse industrial applications.

**Table 1**
Table A: Evaluation table for the upper body segments with RULA scores according to RULA method.

| Table A: | | Wrist | | | | | | | |
|---|---|---|---|---|---|---|---|---|---|
| | | 1 | 1 | 2 | 2 | 3 | 3 | 4 | 4 |
| Arm | Forearm | Wrist turn | | | | | | | |
| | | 1 | 2 | 1 | 2 | 1 | 2 | 1 | 2 |
| 1 | 1 | 1 | 2 | 2 | 2 | 2 | 3 | 3 | 3 |
| 1 | 2 | 1 | 2 | 2 | 2 | 2 | 3 | 3 | 3 |
| 1 | 3 | 2 | 3 | 3 | 3 | 4 | 4 | 4 | 4 |
| 2 | 1 | 2 | 3 | 3 | 3 | 3 | 4 | 4 | 4 |
| 2 | 2 | 3 | 3 | 3 | 3 | 3 | 4 | 4 | 4 |
| 2 | 3 | 3 | 4 | 4 | 4 | 4 | 4 | 5 | 5 |
| 3 | 1 | 3 | 3 | 4 | 4 | 4 | 4 | 5 | 5 |
| 3 | 2 | 3 | 4 | 4 | 4 | 4 | 4 | 5 | 5 |
| 3 | 3 | 4 | 4 | 4 | 4 | 4 | 5 | 5 | 5 |
| 4 | 1 | 4 | 4 | 4 | 4 | 4 | 5 | 5 | 5 |
| 4 | 2 | 4 | 4 | 4 | 4 | 4 | 5 | 5 | 5 |
| 4 | 3 | 4 | 4 | 4 | 5 | 5 | 5 | 6 | 6 |
| 5 | 1 | 5 | 5 | 5 | 5 | 5 | 6 | 6 | 7 |
| 5 | 2 | 5 | 6 | 6 | 6 | 6 | 6 | 7 | 7 |
| 5 | 3 | 6 | 6 | 6 | 7 | 7 | 7 | 7 | 8 |
| 6 | 1 | 7 | 7 | 7 | 7 | 7 | 8 | 8 | 9 |
| 6 | 2 | 8 | 8 | 8 | 8 | 8 | 9 | 9 | 9 |
| 6 | 3 | 9 | 9 | 9 | 9 | 9 | 9 | 9 | 9 |

**Table 2**
Table B: Evaluation table for the neck, trunk, and legs with RULA scores according to RULA method.

| Table B: | | Trunk | | | | | | | | | | | |
|---|---|---|---|---|---|---|---|---|---|---|---|---|---|
| | | 1 | 1 | 2 | 2 | 3 | 3 | 4 | 4 | 5 | 5 | 6 | 6 |
| **Neck** | | Legs | | | | | | | | | | | |
| | | 1 | 2 | 1 | 2 | 1 | 2 | 1 | 2 | 1 | 2 | 1 | 2 |
| **1** | | 1 | 3 | 2 | 3 | 3 | 4 | 5 | 5 | 6 | 6 | 7 | 7 |
| **2** | | 2 | 3 | 2 | 3 | 4 | 5 | 5 | 5 | 6 | 7 | 7 | 7 |
| **3** | | 3 | 3 | 3 | 4 | 4 | 5 | 5 | 6 | 6 | 7 | 7 | 7 |
| **4** | | 5 | 5 | 5 | 6 | 6 | 7 | 7 | 7 | 7 | 7 | 8 | 8 |
| **5** | | 7 | 7 | 7 | 7 | 7 | 8 | 8 | 8 | 8 | 8 | 8 | 8 |
| **6** | | 8 | 8 | 8 | 8 | 8 | 8 | 8 | 9 | 9 | 9 | 9 | 9 |

**Table 3**
Table C: Final RULA score table based on Table A and Table B previous scores.

| Table C: | | Final score | | | | | | | | |
|---|---|---|---|---|---|---|---|---|---|---|
| | | Table B: Neck, trunk and legs | | | | | | | | |
| | | 1 | 2 | 3 | 4 | 5 | 6 | 7 | 8 | 9 |
| Table A: Wrist and arms score | 1 | 1 | 2 | 3 | 3 | 4 | 5 | 5 | 5 | 5 |
| | 2 | 2 | 2 | 3 | 4 | 4 | 5 | 5 | 5 | 5 |
| | 3 | 3 | 3 | 3 | 4 | 4 | 5 | 6 | 6 | 6 |
| | 4 | 3 | 3 | 3 | 4 | 5 | 6 | 6 | 6 | 6 |
| | 5 | 4 | 4 | 4 | 5 | 6 | 7 | 7 | 7 | 7 |
| | 6 | 4 | 4 | 5 | 6 | 6 | 7 | 7 | 7 | 7 |
| | 7 | 5 | 5 | 6 | 6 | 7 | 7 | 7 | 7 | 7 |
| | 8 | 5 | 5 | 6 | 7 | 7 | 7 | 7 | 7 | 7 |
| | 9 | 5 | 5 | 6 | 7 | 7 | 7 | 7 | 7 | 7 |

To enable multimodal and easily reconfigurable standard ergonomic assessments, the tool organizes joint range criteria within a configuration dictionary containing two main keys: "range" and "position". This design enables the system to operate exclusively with instantaneous joint angles, regardless of the motion acquisition method used. Each entry in the dictionary provides specific information for calculating ergonomic scores.

- Under the **"range" key**, the dictionary defines the angle name, its corresponding range, and the primary score assigned. For example, elbow flexion angles within predefined thresholds between 60° and 100° in RULA are assigned a score of 1, while angles outside this range are scored accordingly (McAtamney & Corlett, 1993). This structure enables the code to adjust primary scores based on the defined angular ranges, regardless of whether these ranges correspond to a specific method like RULA or other ergonomic assessment frameworks.
- The **"position" key** in the same dictionary provides additional postural information, specifying how certain joint angles influence the overall ergonomic assessment. Each entry under this key refers to the specific joint angle, the type of adjustment (typically an additional score to be added), and the associated joint value. This mechanism enables score adjustments based on the worker's posture and task-specific conditions.

Integrating this information into a single dictionary simplifies the code configuration and adaptation to different assessment methods. This modular structure enhances flexibility in ergonomic analysis, simplifying adaptation to different assessment methods without modifying the core code. Additionally, it enables new scoring criteria or data sources to be seamlessly integrated, supporting adaptability to various occupational contexts. A representative interface of the final implementation is illustrated in Fig. 2, showcasing the tool's user-centric design and customizable features.

The implemented approach incorporates environmental and force factors manually into the calculation through its interface, based on the specified load and activity type (see Fig. 3). The sensorization of kinetic measurements, while valuable for comprehensive risk assessment, is not yet included in this study as it falls beyond the current research scope.

Building upon the adaptability of the ME-WARD tool, further advancements were made to accommodate data obtained from video-based motion capture systems. The following section 2.2. video *Data Analysis* presents the required modifications to the data processing pipeline, ensuring that joint angles and body postures extracted through computer vision techniques can be seamlessly integrated into the ergonomic computation framework. By leveraging the inherent flexibility of the system, video data inputs were processed with the same scoring methodology, enabling direct comparisons and validations against IMU-based assessments.

*2.2. Video data analysis*

Joint angles and ergonomic assessments were computed from both IMU-based sensor data and video-based recordings using ME-WARD RULA toolkit. The IMU data, exported as CSV files from Xsens MVN

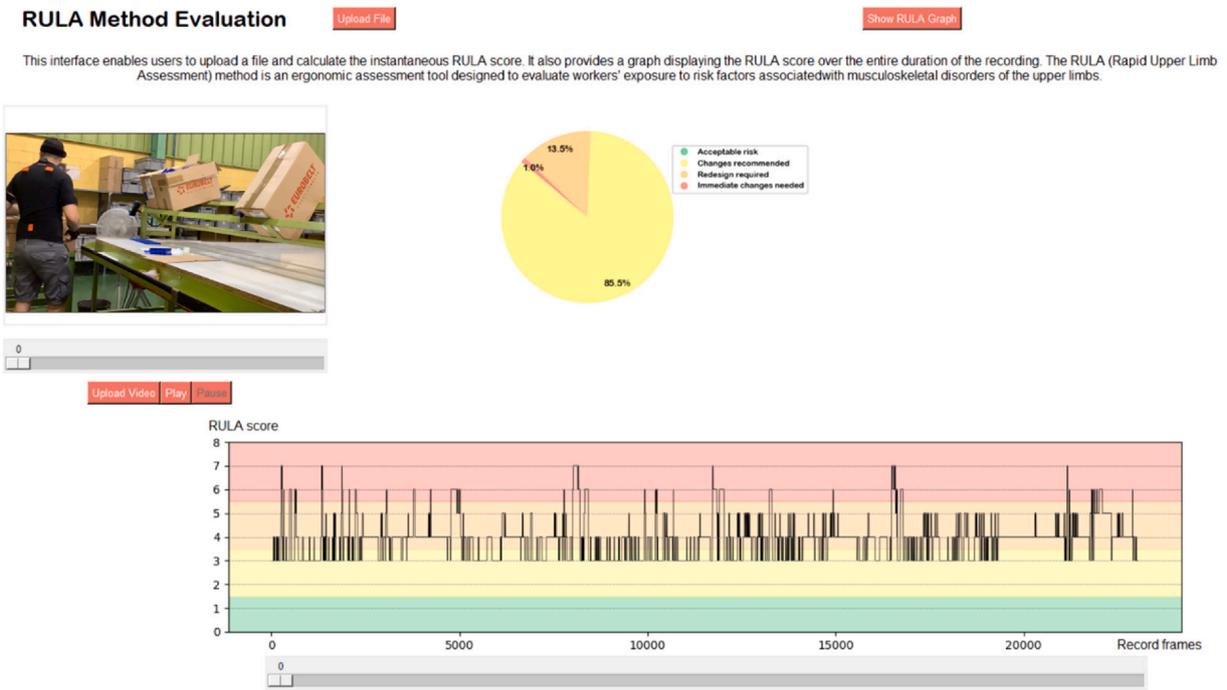

**Fig. 2.** ME-WARD analysis tool interface: resulting video and RULA Graphs.

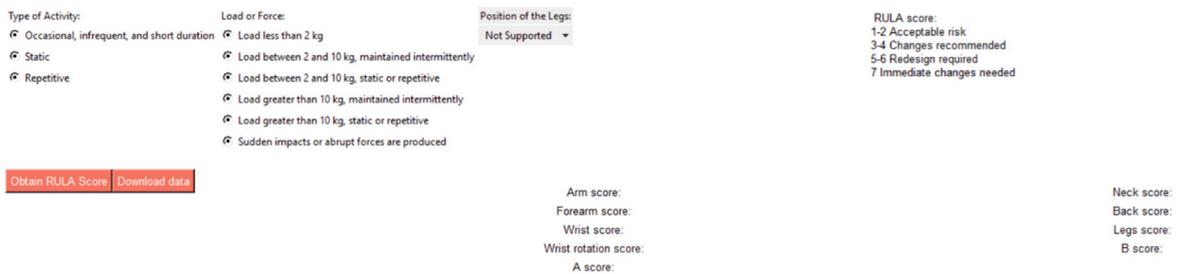

**Fig. 3.** ME-WARD analysis tool interface: RULA additional parameters.

Analyze Pro, consisted of joint angle measurements sampled at 100 Hz. These data were reorganized into time-based intervals for subsequent analysis. The video recordings, captured using an iPad (2022 model), were post-processed with the state-of-the-art pose estimator BodyTrack, included in NVIDIA Maxine-AR-SDK, to extract 3D spatial coordinates of anatomical landmarks, from which joint angles were computed using vector-based methods. A temporal alignment method using minimum Root Mean Square Error (RMSE) was applied to synchronize both datasets, ensuring consistency for direct comparison of joint angles and movement patterns between the two systems. The use of data from NVIDIA Maxine BodyTrack followed the guidelines of the VIDIMU dataset (Martínez-Zarzuela et al., 2023) to obtain key body points from video-based motion capture.

### 2.2.1. Joint angle computation

Joint angles were computed from the 3D spatial coordinates (x, y, z) of key body points by defining two vectors representing adjacent body segments and determining the angular relationship between them using Eq. (1)

$$\theta = \left( \frac{\vec{a} \cdot \vec{b}}{\|\vec{a}\| \|\vec{b}\|} \right) \quad (1)$$

In Eq. (1) $\vec{a}$ and $\vec{b}$ are vectors formed by consecutive body points, "$\cdot$" denotes the dot product, and $\|\vec{a}\|$ and $\|\vec{b}\|$ are the magnitudes of the respective vectors. This equation calculates the angle $\theta$ between the two vectors, corresponding to the joint angle in that plane.

For example, the **elbow flexion** angle is computed using vectors defined by the shoulder-elbow and elbow-wrist points. The computation of each joint angle follows the same approach, leveraging key anatomical landmarks to define vectors representing adjacent body segments:

- **Neck Flexion**: Computed using vectors formed by the torso-neck and neck-nose points, capturing forward or backward head tilt relative to the torso.
- **Neck Rotation**: Defined using the neck and nose points relative to a transverse reference plane, capturing rotational movement of the head around the vertical axis.
- **Neck Bending**: The lateral bending of the neck is determined using the neck and nose points relative to a frontal plane, capturing side-to-side head tilt.
- **Shoulder Flexion**: The arm flexion angle is calculated using vectors formed by the shoulder-elbow and torso-neck points, indicating forward or backward arm movement.
- **Shoulder Adduction/Abduction**: Computed using vectors formed by the shoulder–neck and shoulder-elbow points, capturing the arm's lateral displacement relative to the torso.
- **Shoulder Internal/External Rotation**: Defined using vectors connecting the torso-shoulder and shoulder-elbow points, indicating rotational movement of the shoulder joint.

- **Lumbar Flexion**: Computed using vectors formed by the hips-ankle and pelvis-torso points, capturing flexion or extension of the lumbar spine.
- **Lumbar Rotation**: Computed using vectors formed by the pelvis and hips points in the transverse plane, capturing axial rotation of the lumbar region.
- **Lumbar Bending**: Captures the lateral bending of the lumbar spine using vectors formed by the pelvis and torso points in the frontal plane.
- **Wrist Flexion**: Computed using vectors formed by the elbow-wrist and wrist-finger knuckle points, capturing the flexion and extension of the wrist.
- **Wrist Rotation**: Defined using vectors from the pinky knuckle and wrist-elbow points, capturing twisting motion of the wrist.
- **Wrist Radial/Ulnar Deviation**: Computed using vectors formed by the pinky knuckle and wrist-elbow points, capturing side-to-side wrist movement.

Some keypoints selected for joint angles computation (Fig. 4) were prioritized over alternative landmarks based on their robustness within the deep learning-based body tracking model (NVIDIA Maxine-AR-SDK BodyTrack). For example, in the case of neck angles, while ears could serve as reference points, the nose was selected due to frequent self-occlusion of the left ear. This adjustment minimized errors caused by incomplete visibility. However, using the nose as a reference required an additional step: subtracting the initial inclination of the head (measured at the start of the task) from the final neck flexion angle. This correction ensured an accurate representation of neck movement throughout the task, highlighting the importance of tailoring keypoint selection to address task-specific and environmental challenges.

A function was developed to calculate joint angles from a given set of coordinates. The tool loops through all time frames, ensuring accurate computation of angles throughout the entire task duration. Another function maps these computations to all relevant body joints, including neck flexion, lumbar bending, and wrist rotation. This comprehensive approach enabled the extraction of ergonomic parameters, such as mean joint angles and their variability, providing a robust foundation for statistical analysis.

*2.2.2. Data processing & ergonomic analysis*

After temporal alignment and data conversion, the datasets from both IMU-based and camera-based systems were structured within a Pandas DataFrame, optimized for ergonomic analysis. This setup enabled the computation of:

- Joint angles at specific postures.
- Joint angles over time, enabling comprehensive RULA scoring.

By standardizing joint angle data across both systems, we ensured an objective comparison of ergonomic risk scores, treating the Movella Awinda system as the gold standard for validation.

For both measurement systems, the data processing pipeline facilitated the calculation of:

- Mean joint angles
- Angle variability
- Other relevant ergonomic parameters

These metrics were subsequently used to perform statistical comparisons between the systems, assessing the consistency and reliability of the camera-based approach relative to the IMU-based system.

*2.3. Measurement systems*

In this study, we utilized and compared the RULA metrics obtained with our tool using two different data acquisition systems: the commercial IMU-based Movella Awinda system (Xsens Technologies, 2016) and the monocular 3D pose tracker NVIDIA Maxine AR BodyTrack (NVIDIA, 2021).

*2.3.1. Imu-based System: Movella Awinda*

The IMU-based system, accessed via an MVN Analyze Pro subscription, provides high-precision motion data through sensors strategically placed on key upper-body segments. This system is widely recognized as the gold standard for assessing work-related musculoskeletal disorders (WMSDs) in workplace ergonomics (Huang et al., 2020; Colim et al., 2021). Subject data, including height (175 cm) and foot size (26 cm), were input into the Xsens MVN software during the calibration procedure to ensure accurate motion capture. Additionally, weight (76 kg) and age (36 years) were recorded at the time of testing. The subject provided informed consent prior to participation, acknowledging the study's objectives, procedures, and data handling protocols in compliance with ethical research standards.

Acquisitions were recorded at a sampling rate of 100 Hz using 11 calibrated sensors for upper-body motion analysis, as shown in Fig. 5. The setup involved precise sensor placement following upper-limb sensor placement guidelines (Höglund et al., 2021) and a "walk-around" manufacturer calibration routine to ensure accurate joint tracking throughout the task (Xsens Technologies, 2016). For scenarios where custom or low-cost sensors are used, adherence to the guidelines outlined in González-Alonso et al. (2023) is recommended to maintain consistent and reliable sensor placement, ensuring that the IMU placement in the anatomical model corresponds to real-world sensor placement.

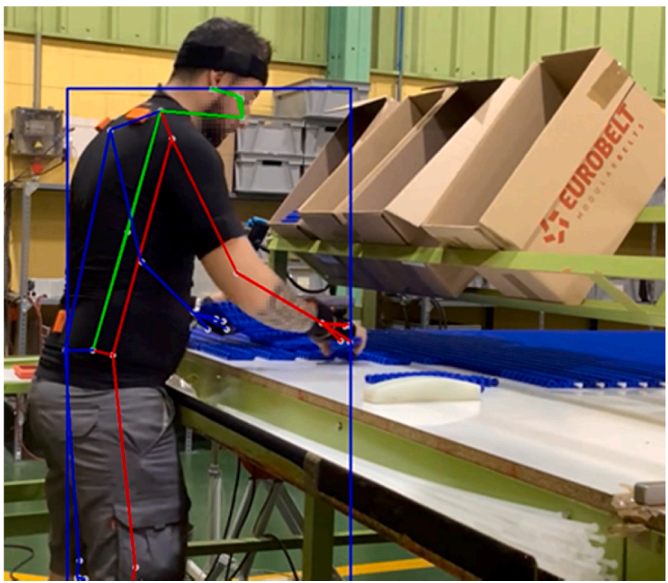

**Fig. 4.** Body keypoints obtained by NVIDIA MaxineAR BodyTrack.

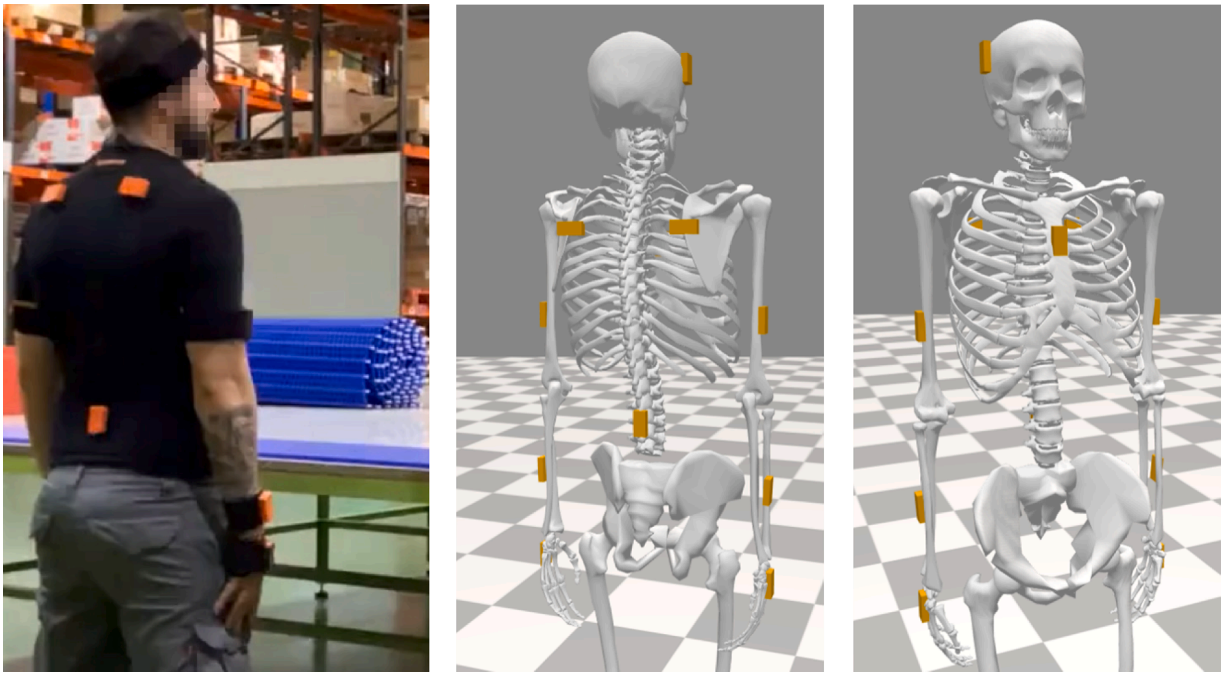

**Fig. 5.** Subject sensors placement: subject on camera, Left −side and Right-side Rajagopal model with sensors on designated locations. The sensors present on the subject and represented in the model are the following: Head (any place that allows the headband); Sternum; RSho: right shoulder, LSho: left shoulder (at the height of the scapulae); RUA: right upper arm, LUA: left upper arm (outer side of the arm); RLA: right lower arm, LLA: left lower arm (before the wrist bone); Rhand: right hand, Lhand: left hand (above the wrist bone); and Hips (located approximately in L4-L5 vertebra).

It is important to acknowledge the potential limitations of IMU-based systems, including sensor drift, shifting, and error accumulation, particularly during long recording sessions and for custom or low-cost sensor systems, as reported by González-Alonso et al. (2021). Prior to introducing recorded data into the ME-WARD tool, necessary precautions should be taken to maintain data integrity. This includes applying appropriate filtering techniques or recalibration procedures, as recommended in the Movella Xsens MTw Awinda User Manual. By proactively addressing these factors, the accuracy and reliability of ergonomic assessments can be maintained across extended data collection periods.

Furthermore, in industrial environments, the potential effects of ferromagnetic interference on IMU sensor measurements must be carefully considered. Ferromagnetic sources, such as machinery and electronic devices, can introduce distortions. Various mitigation techniques have been developed to address these challenges. Commercial systems, such as the Movella Awinda system, incorporate interference correction algorithms within their sensor fusion mechanisms, effectively minimizing the impact of magnetic disturbances (Paulich et al., 2018). Additionally, for custom IMU-based systems, methods such as excluding magnetometer data during high-interference periods have been proposed in the literature (González-Alonso et al., 2021). These strategies help ensure reliable motion tracking even in challenging industrial conditions.

#### 2.3.2. Camera-Based System: NVIDIA Maxine AR BodyTrack

The NVIDIA Maxine AR BodyTrack employed a monocular camera to track key anatomical points in real time, using deep learning for pose estimation. The system was optimized for NVIDIA GPUs with Tensor Cores, making it a cost-effective, non-intrusive solution, such as those found in the NVIDIA Quadro, Tesla, Data Center, or RTX series. This markerless tracking system provides positional data, including joint angles and segment orientations, under appropriate lighting conditions.

To ensure optimal visibility of the worker's joint movements during ergonomic assessment, video recordings were conducted from a semi-sagittal viewpoint rather than a frontal perspective. This decision was driven by practical constraints within the workstation environment, where the presence of conveyor belt systems and assembly components (e.g., large boxes) caused occlusions of critical joint angles when viewed from the front. A rear perspective was also deemed unsuitable due to spatial limitations that prevented complete framing of the worker's movements. The semi-sagittal viewpoint provided an unobstructed view of key upper-limb and trunk movements, ensuring that essential ergonomic parameters could be accurately captured while maintaining alignment with RULA assessment objectives.

During recording, precautions were taken to ensure that the worker remained within the camera's focus throughout the task. Several key parameters were considered to optimize the camera-based system performance, including resolution, frame rate, and field of view. Additionally, it is advisable to dynamically adjust weighting factors based on the reliability of the data for each body segment acquired during different movement phases.

The video recordings were captured using an HD device (iPad 2022) at a resolution of 1280 x 720 pixels (720p), with a frame rate of 30 fps and a video data rate of 4870 kbps. The NVIDIA Maxine AR BodyTrack system was configured accordingly, using a resolution of 1280x720 pixels and a frame rate of 30 fps. The camera positioning was optimized based on the workstation layout to minimize occlusions and maximize visibility of critical body segments, while accommodating the practical constraints of the workplace.

Additionally, IMU and video data were synchronized to ensure both temporal alignment and consistency, enabling accurate comparisons. A common timestamping method was used, combined with predefined calibration movements performed by the worker at the start of each recording session to precisely align the data streams.

The statistical metrics selected for this study, including root mean square error (RMSE) and correlation coefficients, were chosen for their proven effectiveness in evaluating the reliability and consistency of joint angle measurements across different motion capture systems (Choffin et al. 2023; Moghadam et al. 2023; Van Crombrugge et al. 2022). The RMSE provides a quantitative measure of the average difference between the two signals. Correlation coefficients evaluate the strength and direction of the relationship between signals, highlighting the level of agreement in capturing movement patterns. Together, these metrics establish a robust framework for validating multimodal approaches, ensuring that the ME-WARD tool delivers consistent and reliable results across diverse input sources, including both IMU-based and video-based systems.

### 2.4. Acquisition protocol

The acquisition protocol was designed to assess ergonomic risks using IMU-based and monocular systems independently. Data collection followed an ergonomic task flow, recording representative actions such as inserting rods, pushing belt components, and manually selecting parts for both wide and narrow belt tasks. This protocol was specifically tailored to the conveyor belt assembly process and adapted to the constraints of the working environment and task flow.

Before recording, IMUs from Movella Awinda system (Paulich et al., 2018) were placed on key upper-body segments, including the head, scapulae, torso, upper arms, forearms, hands, and hips. The IMU system was calibrated using the manufacturer's calibration process (Xsens Technologies, 2016) to ensure accurate motion capture. Simultaneously, a monocular setup was positioned at a semi-sagittal angle to maximize visibility of right-side movements while minimizing occlusions.

For scenarios where custom or non-commercial IMU sensors are used, it is advisable to perform an initial calibration procedure to ensure accurate data acquisition. A simplified calibration approach, such as the method described by González-Alonso et al. (2024), can be employed to improve the alignment of sensor frames with anatomical body segments. This procedure involves the subject adopting a neutral standing position with arms extended forward (N-pose), which enables the correction of initial sensor misalignment and drift before data collection. Implementing such calibration techniques improves consistency and reliability in joint angle estimations, particularly when using low-cost or custom-built sensor systems.

In our acquisition process, data were recorded continuously throughout the assembly task, with breaks taken as needed to maintain ergonomic comfort and prevent fatigue. The IMU sensor system was placed beneath the worker's Personal Protective Equipment (PPE). Additionally, an adaptation period was incorporated to allow the worker to become familiar with the motion capture equipment (IMUs and camera), reducing discomfort or alterations in natural movement patterns during the recording phase. Multiple capture trials were conducted to ensure that the collected data were representative of actual work conditions, while avoiding interruptions or deviations in the operator's technique.

During data acquisition, the worker performed the conveyor assembly process, beginning by connecting the initial conveyor sections and progressively pushing the assembled parts forward as additional sections were added. At specific intervals, the worker inserted side rods to secure the segments, a critical step to ensure structural stability (Fig. 6). To minimize disruption to the production workflow, the setup and configuration of the acquisition systems were completed beforehand, and the worker was given sufficient time to adjust to wearing the sensors.

Continuous recording intervals were used, with durations long enough to capture the entire assembly process for both wide and narrow conveyor belts, as well as to document the finer steps comprising each task. This ensured comprehensive data collection of the worker's movements throughout the assembly process.

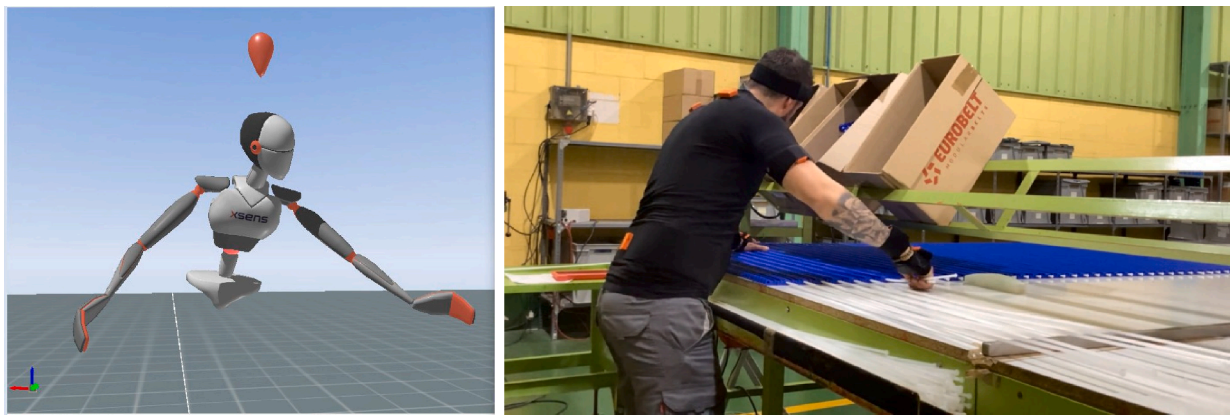

**Fig. 6.** Inserting rods in the wide belt: MVN Analyze (left) and video (right) recordings.

### 2.5. Workstation layout and organization

The workstation under analysis belongs to a manufacturing company specialized in 3D printing of modular plastic conveyor belts for industrial and food sectors. The company is dedicated to assembling conveyor belts, a process carried out within a designated area of the manufacturing facility. The conveyor belt assembly workstations are designed to allow replication of the described layout across multiple stations, enabling parallel assembly of multiple belts as needed (Fig. 4). However, unlike continuous production lines, the assembly tasks at this workstation are customized and specialized. Factors such as space availability, time constraints, and productivity demands are not critical in this setting and, therefore, do not significantly influence the arrangement or efficiency of the workspace. While assembling a belt involves repetitive actions—such as adding modular rows until the desired belt length is achieved—two distinct variations of the recorded task exist: assembling a wide belt and assembling a narrow belt.

The assembly process follows a linear material flow: components are aligned, attached sequentially, and advanced along the workstation until completion. The primary responsibility for assembly lies with the technician, who is supported and overseen by supervisors focused on quality assurance or ergonomic assessments. The technical operator advances the partially assembled belt from the starting position to the end of the workbench until the process is complete.

Materials are transported within the workspace using auxiliary transport systems, such as trolleys. However, data collection for this study focused exclusively on the assembly stages conducted at the workbench. Essential components are stored within reach, with boxes positioned directly in front of the operator and on shelves at specific distances, ensuring efficient workflow and safe handling. Tools such as alignment aiders and rods are strategically placed to minimize unnecessary movement and improve accessibility. To further support worker safety, operators wear PPE to guard against sharp objects. This factor must be considered when placing wearable sensors underneath the PPE.

Finally, the entire recording session was supervised by ergonomists from the Occupational Risk Prevention Department of Eurobelt Valladolid. Their role included overseeing sensor placement and ensuring the proper operation of body-worn units throughout data acquisition. This supervision was crucial to maintaining the reliability of the collected data, reinforcing the validity of the results, and ensuring the overall robustness of the study.

## 3. Results

The collected data were processed, visualized, and analyzed using the custom toolkit developed in this study, enabling the computation of joint angles and corresponding RULA scores for different data acquisition setups. This analysis allowed for a comprehensive assessment of the consistency, reliability, and applicability of the proposed method across various configurations.

The results are presented through a series of evaluations: (1) a comparison between proprietary and custom RULA computation methods, (2) a performance comparison between IMU-based and camera-based systems, and (3) the ergonomic results derived from the RULA scores in the context of workstation assessment.

### 3.1. Proprietary vs custom RULA computation

To validate the effectiveness of the custom RULA computation system, a first comparative analysis was conducted against the proprietary RULA scores provided by Movella's MotionCloud software. Using joint angle data generated by the MVN Pro software, the ME-WARD RULA analysis tool replicated the ergonomic analysis, producing RULA scores that closely aligned with those from the commercial system, as shown in Table 4. This comparative analysis was performed across different belt assembly processes, including picking up and transporting elements on the trolley, the assembly process of the belt itself, attaching cleats to the belts, and their subsequent winding. Fig. 7 illustrate examples of these tasks and their corresponding ergonomic assessments.

RULA scores were selected due to their widespread acceptance in ergonomic risk assessment, providing a standardized method to quantify posture-related musculoskeletal risks. By considering the proprietary RULA scores as the reference standard, this comparison demonstrated the reliability of the developed tool when applied to accurate joint angle data. Consequently, the validated system can be reliably extended to assess whether joint angles obtained from video-based systems provide sufficient information to generate a RULA report equivalent to the one produced with IMUs, ensuring the robustness and applicability of the ergonomic assessment process across diverse motion capture technologies.

### 3.2. Imu-based vs camera-based

The second study compared ergonomic assessments conducted using IMU and monocular systems during conveyor belt assembly tasks. Representative high-risk movements, such as inserting rods, pushing belt components, and manually selecting parts, were simultaneously captured using both systems.

**Table 4**
Movella Awinda MVN Pro license recordings: proprietary Motion Cloud RULA analysis vs ME-WARD RULA analysis tool results on different activities.

| Record | Ergonomic analysis tool | Negligible risk | Low Risk | Medium risk | Very high risk |
|---|---|---|---|---|---|
| Conveyor belt winding | MVN RULA Motion Cloud | 0.0 % | 79.0 % | 13.0 % | 8.0 % |
| | ME-WARD RULA analysis tool | 0.0 % | 78.7 % | 13.4 % | 7.9 % |
| Inserting cleats | MVN RULA Motion Cloud | 0.0 % | 91.0 % | 9.0 % | 0.0 % |
| | ME-WARD RULA analysis tool | 0.0 % | 88.1 % | 11.7 % | 0.2 % |
| Belt Assembly | MVN RULA Motion Cloud | 0.0 % | 65.0 % | 31.0 % | 4.0 % |
| | ME-WARD RULA analysis tool | 0.0 % | 64.0 % | 30.4 % | 5.6 % |
| Picking up and transport | MVN RULA Motion Cloud | 0.0 % | 77.0 % | 22.0 % | 0.0 % |
| | ME-WARD RULA analysis tool | 0.0 % | 76.4 % | 23.4 % | 0.2 % |

**Conveyor belt winding**

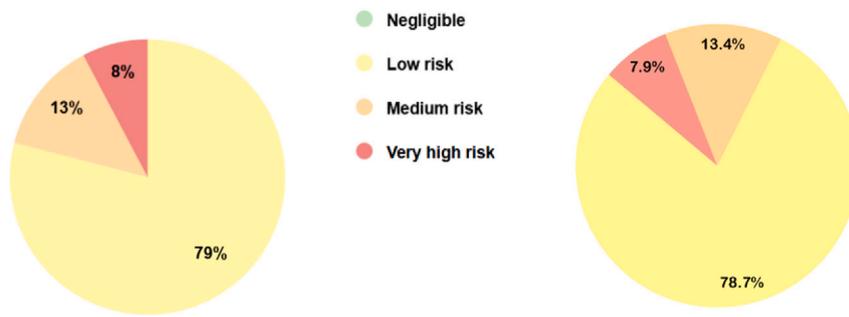

**(a)** RULA pie graph generated with MVN Motion Cloud.      **(b)** RULA pie graph generated with ME-WARD custom tool.

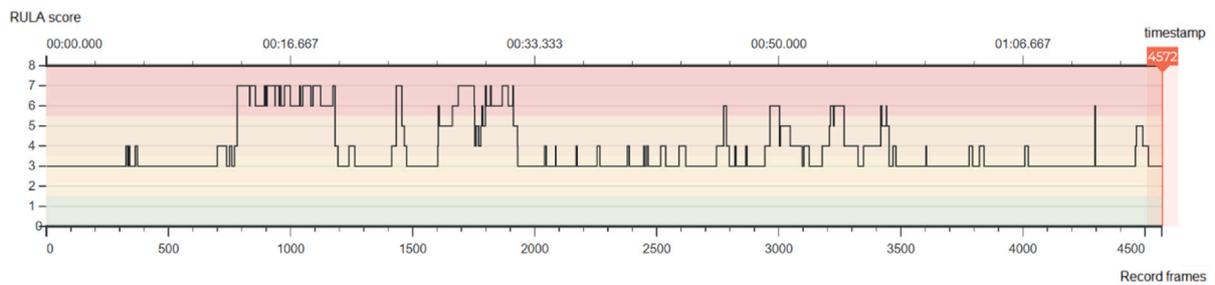

**(c)** RULA score graph generated with MVN Motion Cloud.

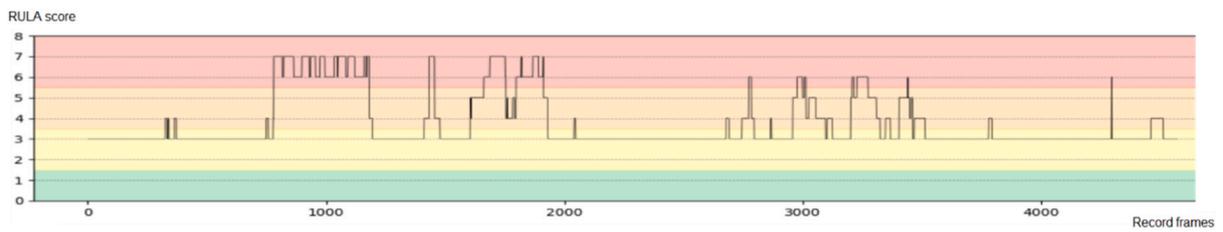

**(d)** RULA score graph generated with ME-WARD custom tool.

**Fig. 7.** Conveyor belt winding recordings: RULA score results.

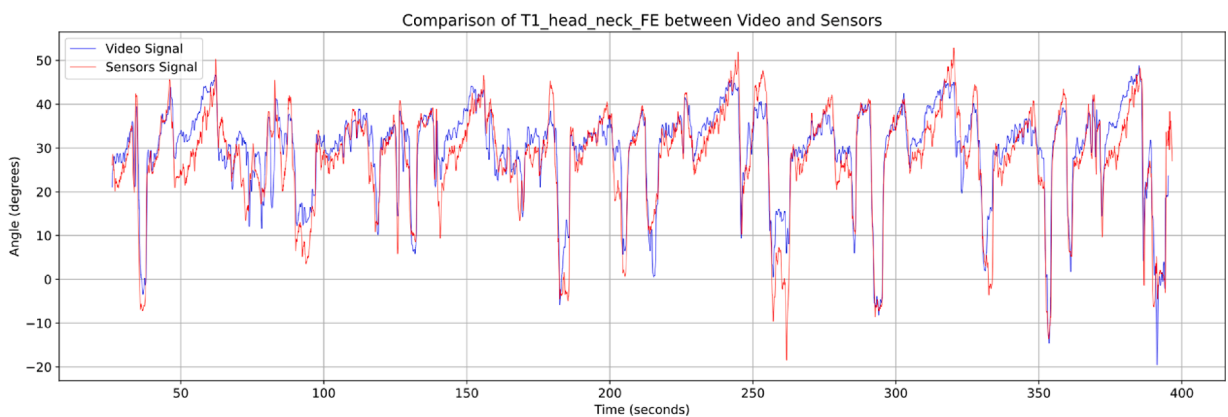

**Fig. 8.** Head neck flexion extension.

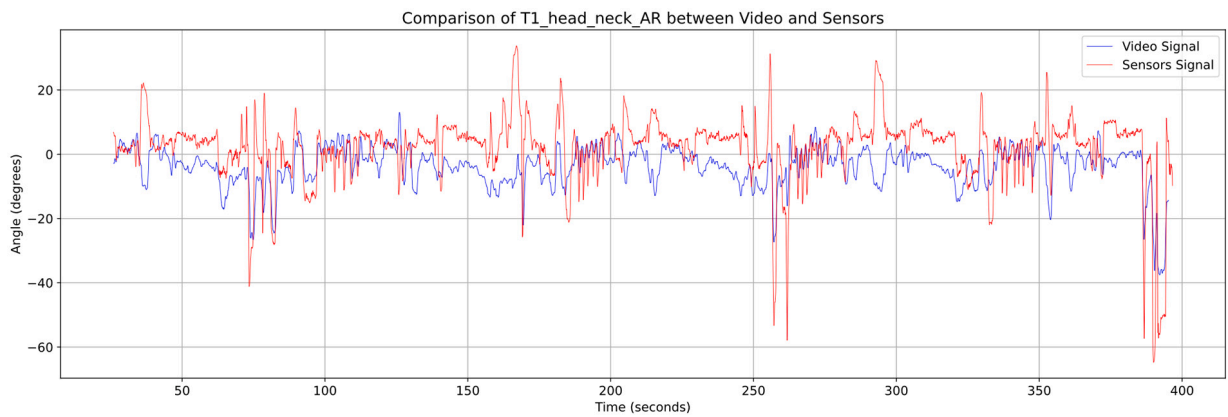

**Fig. 9.** Head neck rotation.

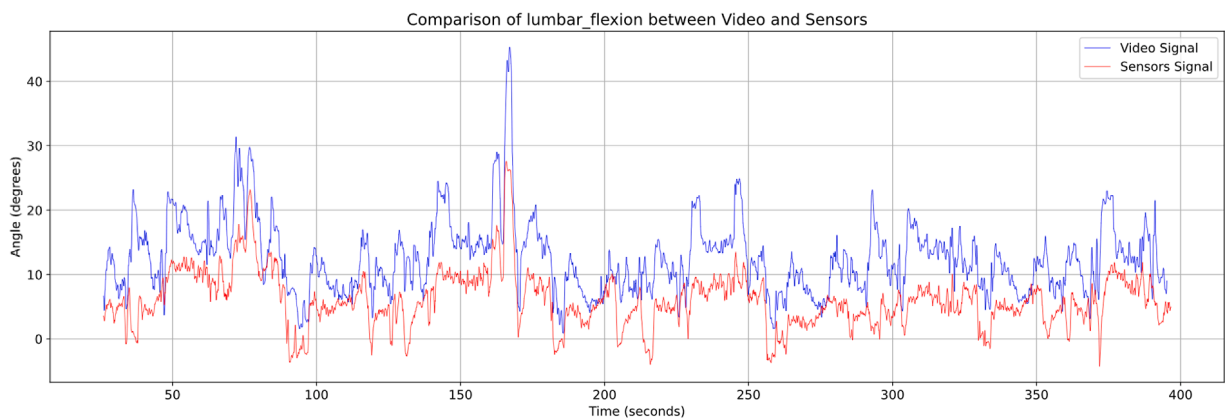

**Fig. 10.** Lumbar flexion extension.

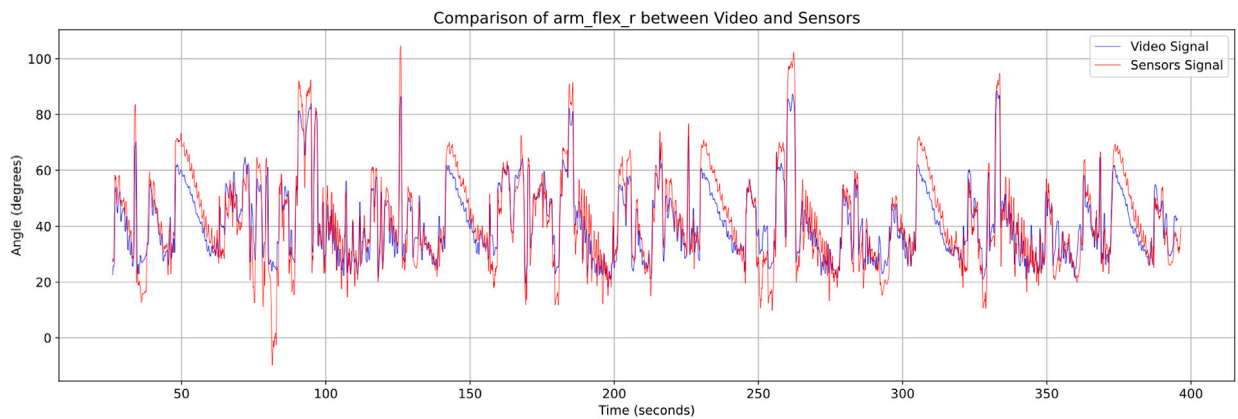

**Fig. 11.** Arm flexion extension right side.

*3.2.1. Joint angles analysis*

The joint angles derived from motion captured with the Movella Awinda sensor system, analyzed using the MVN Analyze Pro license, were compared with the joint angle estimations obtained from the camera-based setup. The Xsens IMU system, extensively validated in prior research as a benchmark for biomechanical assessments (Robert-Lachaine et al., 2017), served as an ideal reference point for evaluating the camera-based methodology.

The IMU system consistently delivered precise joint angle measurements, particularly for lateral and rotational movements. In contrast, the camera-based system excelled in capturing flexion and extension but exhibited limitations in multi-planar tracking due to occlusions or single-view constraints. Figs. 8 to 14 present overlaid graphs illustrating these differences, showing strong alignment in flexion-dominated movements, while noting reduced accuracy for rotations, lateral inclinations, and left-side movements in the camera-based outputs.

**Narrow belt**

As shown in Figs. 8, 11, and 14, the evaluation of neck, shoulder (arm), and elbow flexion revealed that the camera-based system produced results largely consistent with those of the Xsens IMU system for right-side movements. These flexion movements, primarily occurring within a single plane, were effectively captured by the camera system, demonstrating a strong correlation between the two measurement methods. However, as illustrated in Fig. 12, left-side movements exhibit lower performance in the video-based results compared to right-side movements. This discrepancy was observed even in movements such as arm flexion (Fig. 11), where right-side motion closely aligned with sensor-derived values, highlighting a consistent disparity between the two sides.

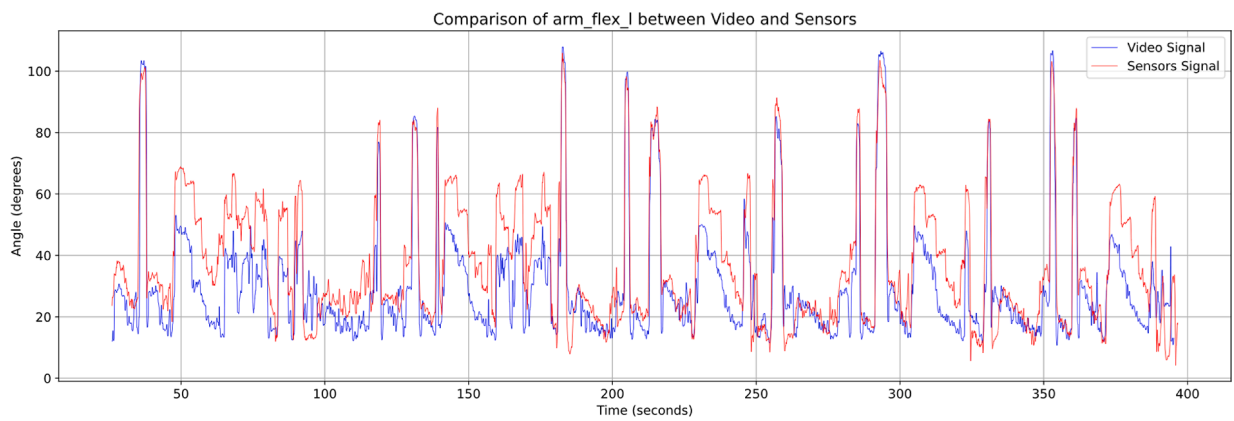

**Fig. 12.** Arm flexion extension left side.

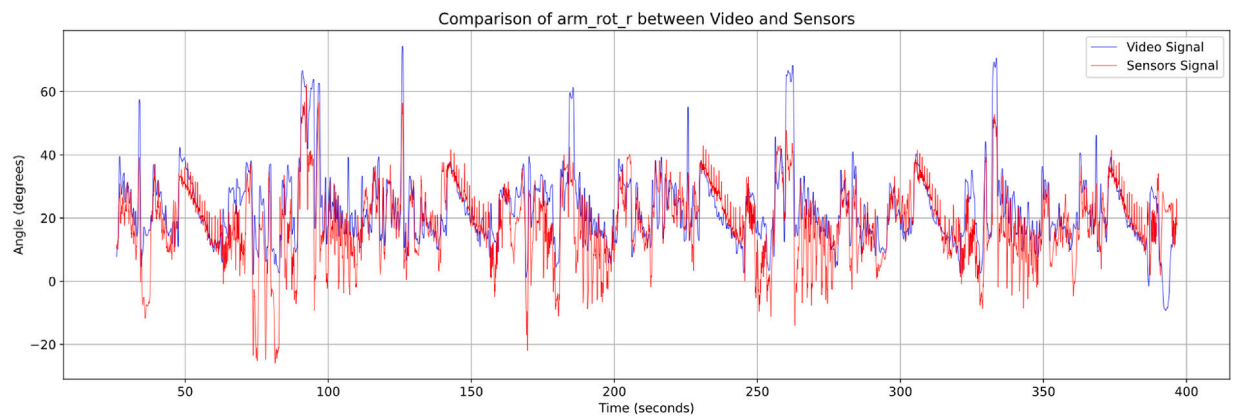

**Fig. 13.** Arm rotation right side.

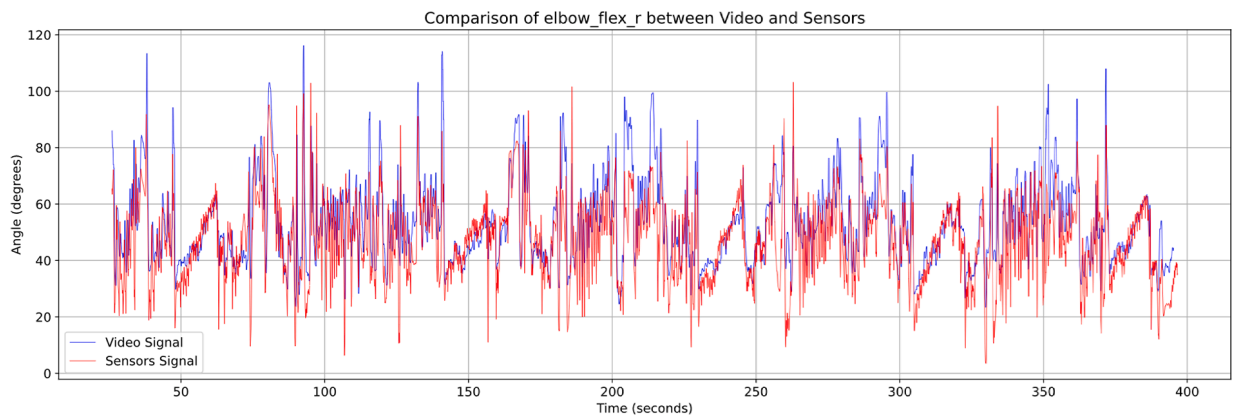

**Fig. 14.** Elbow flexion extension right side.

The camera-based system exhibited notable limitations in accurately capturing joint rotations and lateral inclinations, as illustrated in Figs. 9 and 13. Due to its fixed perspective and reduced depth perception, even with computational adjustments to infer depth, inferred joint angles of these movements from video showed greater discrepancies compared to those obtained from IMU data. The deviations were particularly pronounced in neck and shoulder rotation measurements, highlighting the challenges of relying on a monocular setup for complex three-dimensional joint movements.

**RULA ergonomic results**

Using the multimodal tool, ergonomic risks were quantified for both wide and narrow belt assembly tasks. RULA scores derived from IMU data served as the benchmark, while the camera-based system achieved scores within acceptable ranges for most tasks. As illustrated in Figs. 15 and 16 (subfigures a and b), both systems classified ergonomic risk levels similarly, particularly during high-risk moments. For actions involving repetitive flexion, such as pushing belt segments, the camera-based system generated RULA scores closely matching those of the IMU system, as depicted in the RULA score graphs in Figs. 15 and 16

## Wide Belt

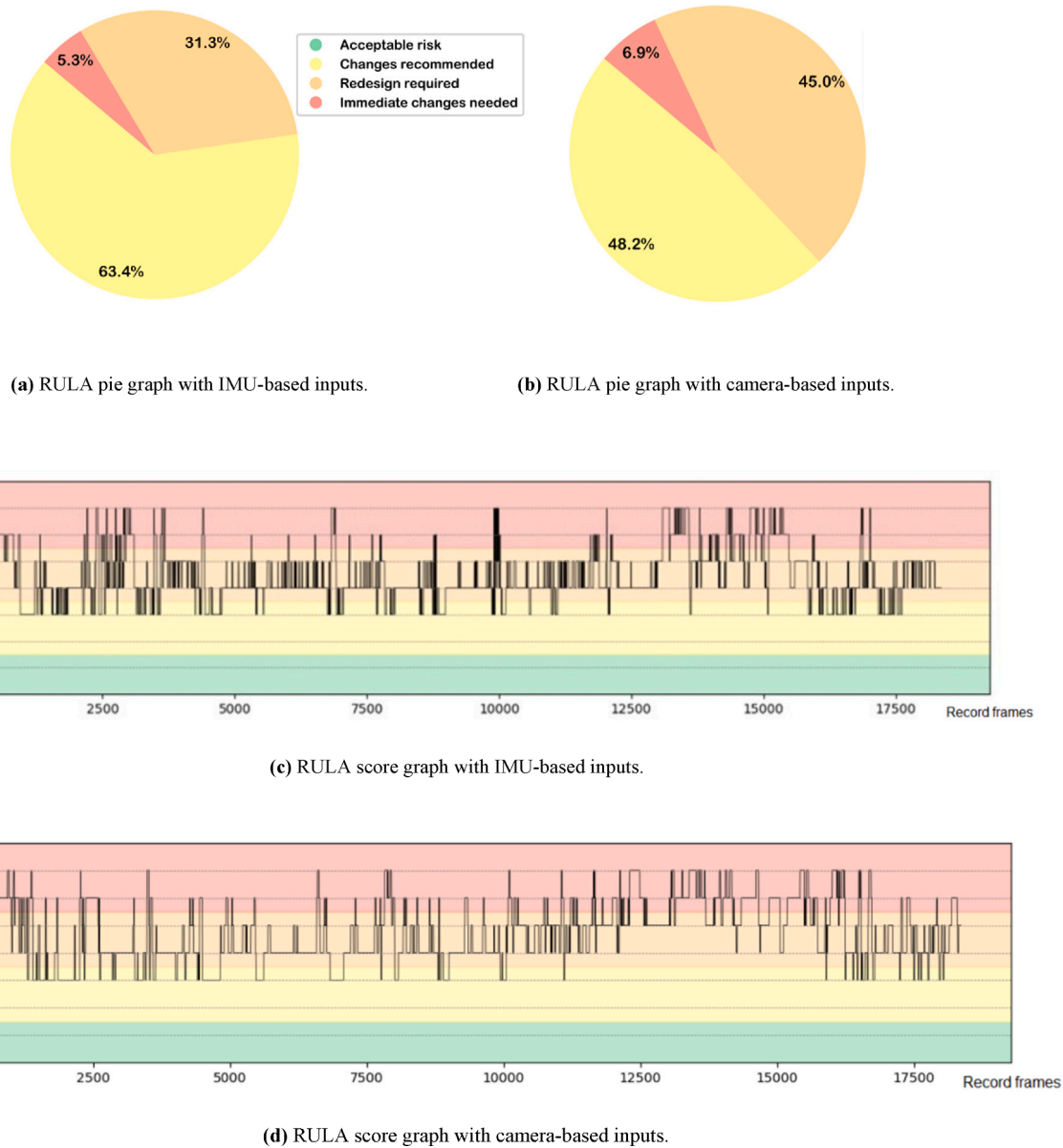

(a) RULA pie graph with IMU-based inputs.

(b) RULA pie graph with camera-based inputs.

(c) RULA score graph with IMU-based inputs.

(d) RULA score graph with camera-based inputs.

**Fig. 15.** Wide Belt RULA pie and score graphs: IMU-based vs. camera-based inputs.

(subfigures c and d). However, for movements requiring precise rotational tracking, slight underestimation of ergonomic risks was observed in the camera-based scores.

### 3.2.2. Statistical analysis

A statistical analysis was conducted to evaluate the correlation and the RMSE between joint angle measurements obtained from IMUs and video-based systems at the two most common workstations: wide belt and narrow belt assembly stations. The RMSE quantifies the deviation between IMU-based and video-based signals, while correlation coefficients indicate the strength of the linear relationships between datasets. These indicators were selected based on their proven effectiveness in previous studies for comparing motion capture technologies and ensuring accurate ergonomic analysis in industrial environments (Van Crombrugge et al. 2022).

Data were collected over three runs per workstation, totaling 40 min of recordings. This analysis provides an in-depth comparison of joint angle measurements captured by both systems, focusing on their agreement across different movement planes. The tables and graphs presented in this section summarize these correlations.

The RMSE was observed to be less than 15 degrees in both wide and narrow belt recordings for primary right-side flexion movements and even for other movements, such as abduction or lateral bending (Tables 5 and 6). The mean of cross-correlation coefficients ranged between 0.66 and 0.9 for flexion-dominated movements, demonstrating strong agreement between the two systems. For multi-planar movements, results exhibited greater variability across different recordings, with correlation coefficients ranging from 0.1 to 0.76, depending on the specific recording conditions. This illustrates the camera-based system limitations in consistently capturing lateral and rotational angles under varying scenarios. The results are presented in Tables 7 and 8, further illustrating the robustness of the tool for general ergonomic assessments. However, the statistical results for lateral and rotational movements indicate weaker correlations. Wrist movements also yielded poor results

**Narrow Belt**

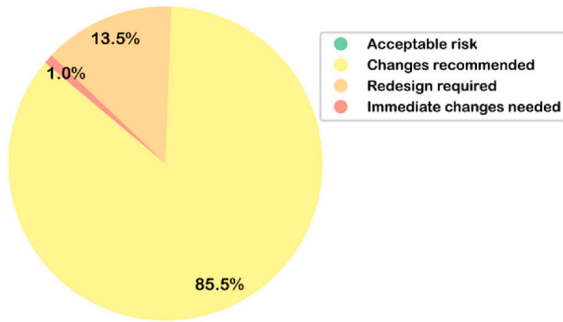
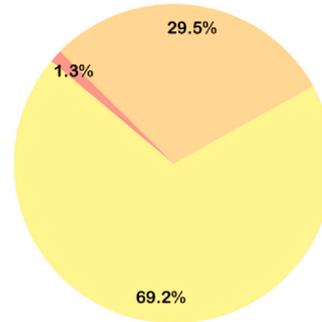

(a) RULA pie graph with IMU-based inputs.     (b) RULA pie graph with camera-based inputs.

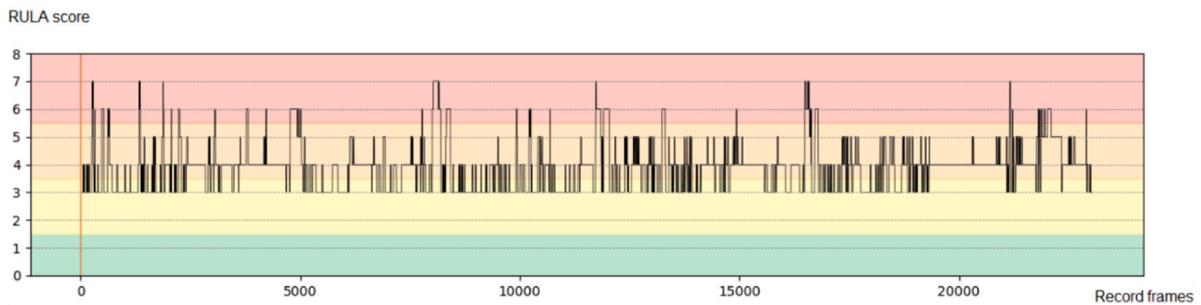

(c) RULA score graph with IMU-based inputs.

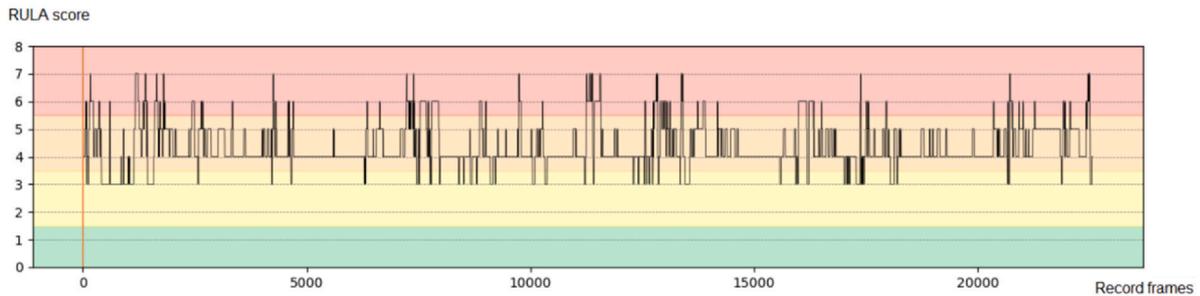

(d) RULA score graph with camera-based inputs.

**Fig. 16.** Narrow Belt RULA pie and score graphs: IMU-based vs. camera-based inputs.

and were excluded from the analysis, as their correlation values were below 0.3 in most cases. An example of this correlation analysis from wide and narrow belt recordings is shown in Figs. 17 and 18. Despite these limitations, the camera-based system still demonstrates acceptable performance for simpler movements, as detailed in the corresponding tables and charts.

### 4. Discussion

In this study we describe and validate ME-WARD, an ergonomic assessment tool capable of processing data from both IMU-based and camera-based systems. The results demonstrate that both approaches to human body movement acquisition can serve as inputs for ergonomic assessment within a unified multimodal framework. This validation underscores the tool's versatility, enabling the digitalization of RULA across diverse motion capture setups.

Although this study did not explicitly measure the accuracy of the IMU system (Movella Awinda), its established reliability in prior research and its consistent performance across all movement types provide a strong foundation for its use as a gold standard reference. Using Xsens IMU data as a benchmark, the comparative analysis suggests that IMUs may offer more consistent measurements due to their ability to capture motion across multiple axes, potentially improving reliability in joint angle estimation.

On the other hand, RULA assessments using the video-based approach exhibited accuracy variations depending on worker self-occlusions and the specific plane of limb movement (flexion, abduction, rotation). In our study, movement acquisition was sufficiently robust for RULA assessments, particularly for flexion movements, which were the most critical in the conveyor assembly process evaluated. However, the camera-based system demonstrated limitations in tracking lateral inclinations and rotational movements, where its monocular setup resulted in reduced accuracy. The observed variability in wrist and trunk movements can be attributed to the camera's monocular perspective limitations, where depth perception plays a crucial role.

Statistical analysis of joint angle measurements revealed high

**Table 5**
Wide belt: RMSE per joint angle between IMU-based and camera-based systems.

| | Wide Belt Recordings | | | |
|---|---|---|---|---|
| RMSE | **Record 1** | **Record 2** | **Record 3** | **MEAN** |
| T1_head_neck_AR | **12.734** | **14.676** | **16.822** | **14.744** |
| T1_head_neck_FE | **8.962** | **11.131** | **11.548** | **10.547** |
| T1_head_neck_LB | **12.316** | **13.757** | **13.777** | **13.283** |
| arm_add_l | 80.108 | 84.718 | 83.948 | 82.924 |
| arm_add_r | **12.704** | **12.814** | **13.310** | **12.943** |
| arm_flex_l | 18.176 | 20.233 | 18.001 | 18.803 |
| arm_flex_r | **9.137** | **11.592** | **12.594** | **11.108** |
| arm_rot_l | 57.726 | 55.076 | 56.463 | 56.422 |
| arm_rot_r | 16.658 | 19.028 | 20.294 | 18.660 |
| elbow_flex_l | 21.455 | 22.091 | 24.036 | 22.527 |
| elbow_flex_r | **12.244** | **15.820** | **18.764** | **15.609** |
| lumbar_bending | **6.845** | **7.025** | **6.763** | **6.877** |
| lumbar_flexion | **6.872** | **5.483** | **5.529** | **5.961** |
| lumbar_rotation | 20.270 | 27.096 | 26.694 | 24.687 |
| pro_sup_l | 38.808 | 39.913 | 41.214 | 39.978 |
| pro_sup_r | 25.211 | 22.924 | 23.262 | 23.799 |
| wrist_dev_l | 31.662 | 33.733 | 34.313 | 33.236 |
| wrist_dev_r | 14.100 | 13.288 | 14.346 | 13.911 |
| wrist_flex_l | 45.669 | 46.448 | 46.211 | 46.109 |
| wrist_flex_r | 14.905 | 16.479 | 15.265 | 15.550 |

**Table 6**
Narrow belt: RMSE per joint angle between IMU-based and camera-based systems.

| | Narrow Belt Recordings | | | |
|---|---|---|---|---|
| RMSE | **Record 1** | **Record 2** | **Record 3** | **MEAN** |
| T1_head_neck_AR | **12.446** | **11.198** | **12.121** | **11.922** |
| T1_head_neck_FE | **9.171** | **5.446** | **8.550** | **7.722** |
| T1_head_neck_LB | **9.407** | **8.029** | **10.491** | **9.309** |
| arm_add_l | 88.051 | 87.904 | 88.721 | 88.225 |
| arm_add_r | **13.673** | **14.423** | **14.850** | **14.315** |
| arm_flex_l | 15.158 | 12.232 | 11.142 | 12.844 |
| arm_flex_r | **11.047** | **6.787** | **7.149** | **8.328** |
| arm_rot_l | 55.265 | 55.407 | 54.966 | 55.213 |
| arm_rot_r | **15.247** | **10.585** | **11.291** | **12.374** |
| elbow_flex_l | 22.534 | 19.824 | 21.829 | 21.396 |
| elbow_flex_r | **11.743** | **12.204** | **13.317** | **12.421** |
| lumbar_bending | 6.654 | 6.363 | 6.885 | 6.634 |
| lumbar_flexion | **8.516** | **7.556** | **7.055** | **7.709** |
| lumbar_rotation | 17.024 | 15.066 | 17.912 | 16.667 |
| pro_sup_l | 34.766 | 38.033 | 35.350 | 36.050 |
| pro_sup_r | 23.109 | 22.481 | 22.727 | 22.772 |
| wrist_dev_l | 35.470 | 32.642 | 31.973 | 33.362 |
| wrist_dev_r | 14.647 | 14.612 | 15.638 | 14.966 |
| wrist_flex_l | 48.850 | 52.410 | 51.732 | 50.997 |
| wrist_flex_r | 15.820 | 14.178 | 15.048 | 15.015 |

**Table 7**
Wide belt: Cross-correlation per joint angle between IMU-based and camera-based systems.

| | Wide Belt Recordings | | | |
|---|---|---|---|---|
| Cross Corr. | **Record 1** | **Record 2** | **Record 3** | **MEAN** |
| T1_head_neck_FE | **0.77** | **0.69** | **0.70** | **0.72** |
| T1_head_neck_AR | **0.76** | **0.69** | **0.67** | **0.71** |
| T1_head_neck_LB | 0.38 | 0.46 | 0.49 | 0.44 |
| lumbar_flexion | **0.70** | **0.70** | **0.76** | **0.72** |
| lumbar_rotation | 0.43 | 0.28 | 0.18 | 0.30 |
| lumbar_bending | 0.33 | 0.35 | 0.34 | 0.34 |
| arm_flex_r | **0.88** | **0.80** | **0.79** | **0.82** |
| arm_flex_l | **0.82** | **0.74** | **0.77** | **0.78** |
| arm_add_r | **0.65** | **0.69** | **0.67** | **0.67** |
| arm_add_l | 0.26 | 0.30 | 0.11 | 0.22 |
| arm_rot_r | 0.67 | 0.58 | 0.61 | 0.62 |
| arm_rot_l | 0.48 | 0.13 | 0.26 | 0.29 |
| elbow_flex_r | **0.77** | **0.63** | **0.56** | **0.66** |
| elbow_flex_l | 0.43 | 0.56 | 0.52 | 0.50 |

**Table 8**
Narrow belt: Cross-correlation per joint angle between IMU-based and camera-based systems.

| | Narrow Belt Recordings | | | |
|---|---|---|---|---|
| Cross Corr. | **Record 1** | **Record 2** | **Record 3** | **MEAN** |
| T1_head_neck_FE | **0.74** | **0.92** | **0.79** | **0.82** |
| T1_head_neck_AR | 0.52 | 0.59 | 0.54 | 0.55 |
| T1_head_neck_LB | 0.40 | 0.12 | 0.26 | 0.26 |
| lumbar_flexion | **0.79** | **0.69** | **0.68** | **0.72** |
| lumbar_rotation | 0.30 | 0.56 | 0.29 | 0.38 |
| lumbar_bending | 0.45 | 0.58 | 0.27 | 0.43 |
| arm_flex_r | **0.88** | **0.92** | **0.90** | **0.90** |
| arm_flex_l | **0.78** | **0.88** | **0.88** | **0.85** |
| arm_add_r | 0.30 | 0.18 | 0.34 | 0.27 |
| arm_add_l | 0.33 | 0.27 | 0.29 | 0.30 |
| arm_rot_r | **0.68** | **0.64** | **0.66** | **0.66** |
| arm_rot_l | 0.08 | 0.49 | 0.33 | 0.30 |
| elbow_flex_r | **0.79** | **0.75** | **0.74** | **0.76** |
| elbow_flex_l | 0.45 | 0.49 | 0.38 | 0.44 |

Note: Article title copied from Manuscript pdf and Figures above title deleted kindly check.

correlation coefficients (r = 0.66–0.92) for flexion movements, indicating the system reliability in capturing key ergonomic metrics. However, lower and more variable correlations (r = 0.1–0.76) for lateral and rotational angles underscore the inherent challenges of monocular systems in handling complex, multi-angular dynamics. A deeper analysis of the error distribution in lateral and rotational movements reveals that discrepancies primarily occur at extreme ranges of motion, where monocular camera perspective effects and self-occlusions impact accuracy. These findings suggest the need for post-processing corrections or the integration of supplementary data sources to minimize tracking deviations.

To address these challenges, future implementations could integrate supplementary data sources from depth cameras or multi-camera setups to improve tracking accuracy. For example, a multi-camera configuration, with cameras positioned laterally, above (zenith view), and behind the worker, could enhance performance, as recommended by Pagnon et al. in their "Pose2Sim" workflow for 3D markerless kinematics (Pagnon et al., 2022). Alternatively, hybrid systems combining IMUs with video data may further optimize accuracy for tasks requiring comprehensive tracking.

Despite the limitations of our monocular study, the strong statistical alignment between the camera-based and reference systems supports the feasibility of camera-based methods as cost-effective alternatives for ergonomic risk assessments. These systems are particularly advantageous in scenarios where workers cannot wear sensors or in resource-constrained environments. The multimodal tool successfully computed RULA ergonomic risk scores from both IMU-based and camera-based inputs. For flexion-heavy tasks such as inserting rods and pushing conveyor segments, RULA scores derived from both systems were closely aligned, confirming the tool's reliability in these scenarios. Furthermore, these findings expand the applicability of methods like RULA, making them more accessible for diverse occupational scenarios and facilitating broader adoption of ergonomic assessment tools. However, reduced accuracy in lateral and rotational angles slightly affected the camera-based system performance in tasks requiring more complex postural tracking.

Although our study does not include the direct measurement of kinetic data, the ME-WARD tool enables manual input of these data at predefined intervals during the recording, as shown in the checkboxes in Fig. 3. Factors such as muscle use frequency and exerted forces are currently assessed through these manual inputs, following the approach used in the compared commercial system Xsens Awinda MotionCloud. Nevertheless, future iterations of our tool could incorporate kinetic data estimation through the OpenSim JointReaction analysis, which estimates joint reaction forces using biomechanical modeling, as described by Seth et al. (2018). This enhancement would provide a more comprehensive understanding of the physical demands across different tasks. Additionally, postural load could be automatically quantified by defining static postures as joint angles that do not vary more than a percentage (e.g., 20 %) of the joint range of motion from a maintained position. Similarly, job frequency could be determined by counting job cycle repetitions within a set period and assessing their duration. These capabilities would be particularly useful for extending the tool to incorporate other upper body ergonomic assessment methods widely used in industry, such as the Ovako Workplace Posture Assessment System (OWAS) (Karhu, Kansi, & Kuorinka, 1977) or the Rapid Entire Body Assessment (REBA) (Hignett & McAtamney, 2000). Even though collecting kinetic data in dynamic industrial environments presents challenges—such as force measurement complexities and the need for precise calibration—integrating these elements into future versions of ME-WARD will enhance the accuracy and depth of ergonomic assessments.

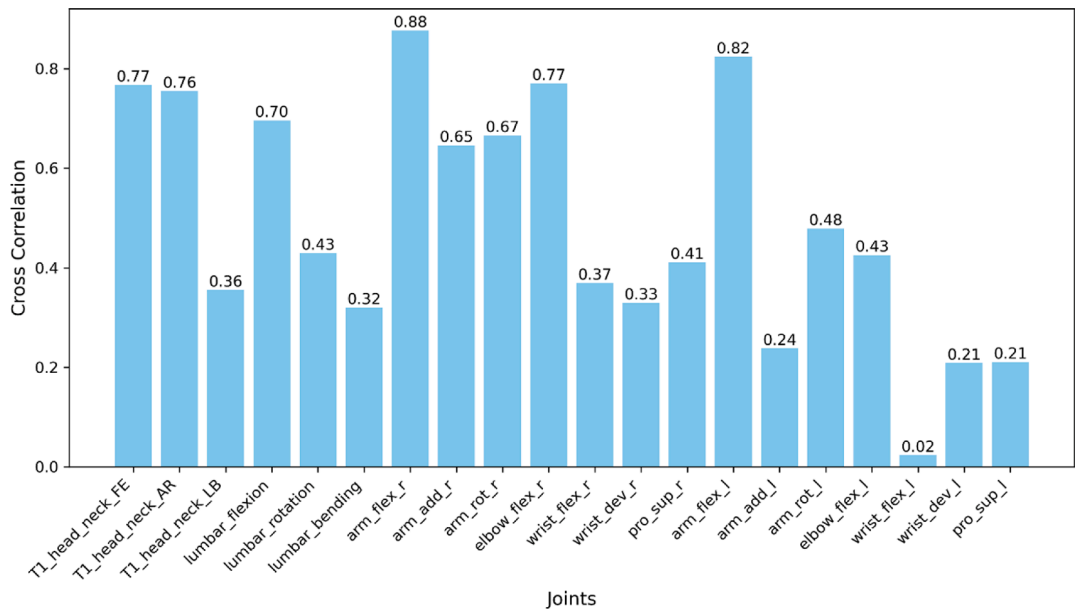

**Fig. 17.** Wide belt: cross-correlation graph per joint angle between IMU-based and camera-based systems.

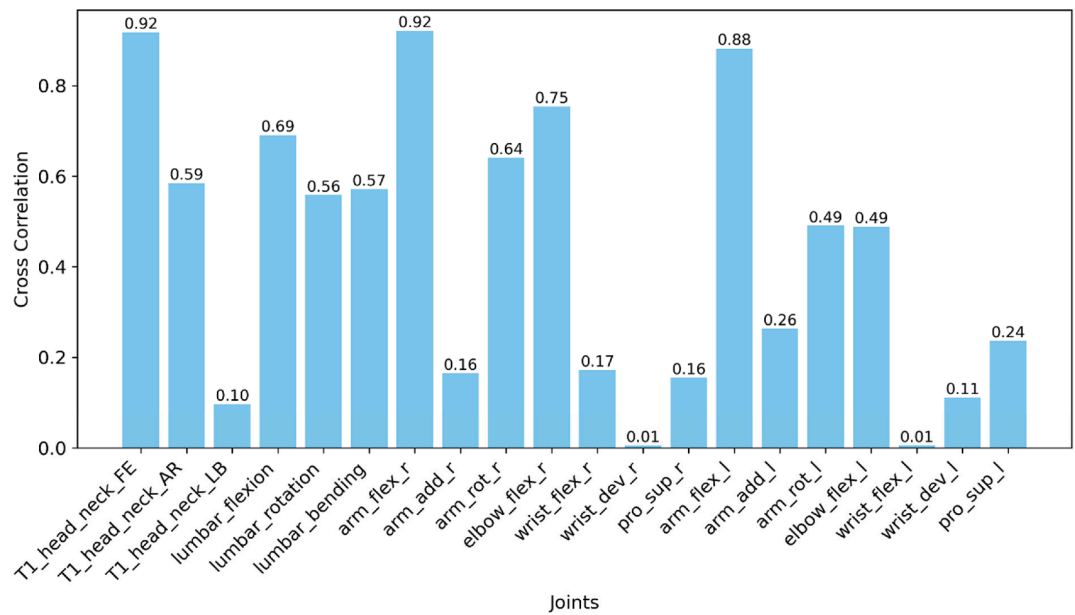

**Fig. 18.** Narrow belt: cross-correlation graph per joint angle between IMU-based and camera-based systems.

Unlike manual ergonomic assessments, such as Nayak et al. (2021), which rely on subjective expert observations, ME-WARD automates the RULA scoring process using standardized algorithms, enhancing reproducibility and minimizing human error. Prior IMU- and camera-based studies validated REBA and RULA scores obtained from these technologies through expert assessment, lacking quantitative accuracy metrics. ME-WARD addresses this by incorporating RMSE and correlation analyses against an IMU-based gold standard, ensuring statistical rigor. Additionally, while some methods require manual post-processing, ME-WARD automates data integration for effective workplace monitoring.

By supporting both IMU-based and video-based motion capture as input, ME-WARD leverages the strengths of each modality for ergonomic risk assessment in diverse industrial settings. Conversely, previous studies, such as Maurer-Grubinger et al. (2021), Zelck et al. (2022), and Baklouti et al. (2024), have focused exclusively on IMU-based systems, achieving reliable results but encountering specific challenges. For instance, Zelck et al. (2022), while using mid-range commercial IMU systems such as Wearnotch, reported sensor drift and electromagnetic interference in port environments, which affected tracking accuracy. Similarly, IMU-based motion capture, while offering high accuracy (Caputo et al., 2019), requires extensive calibration and may be sensitive to drift, which must be considered for rapid industrial deployment. While IMU technology is valuable for motion analysis, supporting both IMU and camera-based motion capture, as in ME-WARD, allows for greater flexibility and adaptability in dynamic workplaces. Moreover, this multimodal approach enables the use of cost-effective IMU solutions, such as those proposed by Gonzalez-Alonso et al. (2021), which incorporate methods to mitigate ferromagnetic disturbances.

Significantly, IMU-based studies have primarily been conducted in controlled environments, such as dentistry (Maurer-Grubinger et al., 2021), container lashing (Zelck et al., 2022), and cable manufacturing (Baklouti et al., 2024), which may limit their generalizability to broader workplace conditions. This limitation is even stronger in traditional multi-camera systems making marker-based tracking impractical for real-world workplace ergonomics (Panariello et al., 2022). ME-WARD, however, is designed as a scalable, multimodal solution adaptable to a wide range of industrial applications.

In contrast, deep learning markerless pose trackers that can use regular cameras are inherently more cost-effective, offer faster setup, and are easier to use, making them an attractive alternative for many industrial scenarios. In this regard, deep learning-based RULA assessment methods, such as those by L. Li et al. (2019), Nayak et al. (2021), and H. Li et al. (2024), have demonstrated the potential of video-based motion capture for ergonomic evaluation. These approaches achieve reliable posture assessments, yet they face certain challenges. L. Li et al. (2019) and Nayak et al. (2021) use 2D pose estimation, which performs well in controlled environments but is susceptible to projection errors when the camera is not strictly frontal, leading to greater inaccuracies in joint angle estimation. Similarly, H. Li et al. (2024)'s Kinect-based system provides 3D skeletal tracking but is constrained by occlusions and fixed camera positioning, which limit its use in unstructured industrial settings. As noted in the study by Manghisi et al. (2017), effective implementation of camera-based systems depends on optimizing workstation layouts to minimize occlusions and enhance data accuracy. Training data generalization is another challenge. Deep learning-based methods such as L. Li et al. (2019) and Nayak et al. (2021) are often trained on static datasets (e.g., Human3.6 M), effectively capturing predefined movements but struggling to generalize to dynamic industrial settings. Additionally, while H. Li et al. (2024) integrates object weight estimation via Faster R-CNN, this method relies solely on visual inference, which is less reliable for dynamic load handling, particularly in scenarios with occlusions or varying lighting conditions.

Future studies could explore the integration of video recordings to refine initial IMU calibration procedures and evaluate the effectiveness of hybrid systems. Such an approach could mitigate the limitations of monocular setups while preserving the accessibility and practicality of video-based methods. Addressing these challenges could further enhance the scalability and versatility of ergonomic assessments, making them more accessible across industries and contributing to occupational health and safety—particularly in settings with budgetary or logistical constraints.

The modular architecture of ME-WARD enables flexibility in integrating additional ergonomic assessment frameworks, ensuring adaptability to different contexts. The system enables customizable sensor placement, calibration settings, and adaptable scoring methods to suit different workplace conditions. The modular design also supports the integration of additional data sources, such as force sensors or electromyography (EMG), or the addition of other camera sources, to provide a more holistic ergonomic assessment. This scalability enables customization for specific operational needs, making it a versatile, future-proof solution for ergonomic risk assessment. Future developments will explore expanding the system to support a wider range of occupational tasks, enhancing its usability across diverse industries with minimal configuration efforts.

## 5. Conclusion

This study evaluated the feasibility and reliability of a multimodal digital ergonomic assessment tool capable of processing data from both IMU-based and camera-based motion capture systems. The results demonstrate the tool's adaptability to diverse input methods, offering a flexible and scalable solution for ergonomic assessments across various workplace environments.

The findings confirm that both systems effectively support ergonomic risk assessments, with each offering distinct strengths and limitations. IMU-based systems provide robust and reliable data for calculating RULA scores, particularly for multi-planar and rotational movements. Their ability to consistently capture complex motion patterns, unaffected by environmental factors such as occlusions, makes them highly reliable. Despite these promising results, several limitations must be acknowledged. The accuracy of RULA scores depends on proper sensor placement and calibration. Although guidelines for correct sensor placement have been provided, they must be carefully followed in non-commercial systems to ensure accurate results. Furthermore, long-duration use in highly dynamic environments may pose challenges related to data drift and system robustness, especially when using customized sensors that lack built-in correction mechanisms in their sensor fusion implementation. These considerations highlight the need for further refinement and validation before large-scale industrial adoption.

Conversely, the camera-based system, utilizing the NVIDIA Maxine model, provided a cost-effective and accessible alternative for tasks primarily involving flexion movements, such as those of the shoulders, neck, and elbows. Its ease of use and rapid setup make it advantageous in settings where IMUs may be impractical due to cost, maintenance, or operational constraints. However, its monocular setup introduced notable limitations, particularly in capturing lateral inclinations and rotational dynamics, leading to reduced accuracy in multi-plane movements. These findings suggest that a hybrid approach integrating video and a reduced number of IMU sensors could enhance overall assessment accuracy by leveraging the strengths of each system.

The developed multimodal tool represents a significant advancement in ergonomic assessment by allowing the incorporation of diverse motion capture technologies within a unified framework. By accurately calculating inter-segment angular variations from camera data, the system bridges the gap between traditional IMU-driven approaches and accessible, non-invasive alternatives. This approach enhances the accessibility of RULA assessments while maintaining the accuracy needed to mitigate ergonomic risks effectively in flexion-heavy tasks.

Future research directions should focus on several key areas: optimizing camera-based system performance for complex joint dynamics, which may involve multi-camera configurations or advanced computer vision models (e.g., MotionBERT presented in Zhu et al., 2023) to address lateral and rotational tracking challenges; combining multiple video models simultaneously or integrating hybrid video-sensor systems to improve accuracy and robustness in motion analysis; enhancing the system capability to incorporate kinetic measurements through biomechanical modeling techniques, to provide a more comprehensive understanding of physical demands; or quantifying exertion exposure and postural load using IMU-based automatic detection, enabling real-time tracking of job cycle repetitions and static posture durations.

Furthermore, future studies should include longitudinal studies to evaluate the long-term impact of the ME-WARD tool on reducing ergonomic risks and improving workplace safety across diverse industrial settings. Expanding the tool's capabilities to support other ergonomic frameworks, such as REBA and OWAS, would ensure broader applicability and compliance with industry standards. Additionally, improving usability based on worker feedback will also be explored to facilitate adoption and ease of use in real-world environments.

Finally, by introducing a versatile, data-driven framework for ergonomic analysis, this study contributes to making RULA assessments more accessible and practical. Continued refinement of the proposed approach has the potential to revolutionize ergonomic risk assessment, offering cost-effective and accurate solutions for diverse industrial applications.

**Informed Consent Statement**

Informed consent was obtained from the worker involved in the study.

**Declaration of generative AI and AI-assisted technologies in the writing process**

During the preparation of this work, the authors used [ChatGPT/DeepL] during the revision process in order to improve grammar and readability. After using this tool/service, the authors reviewed and edited the content as needed and take full responsibility for the content of the publication

**CRediT authorship contribution statement**

**Javier González-Alonso:** Conceptualization, Data curation, Formal analysis, Investigation, Software, Visualization, Writing – original draft. **Paula Martín-Tapia:** Investigation, Software. **David González-Ortega:** Resources, Investigation, Writing – review & editing. **Míriam Antón-Rodríguez:** Resources, Investigation, Writing – review & editing. **Francisco Javier Díaz-Pernas:** Resources, Supervision, Writing – review & editing. **Mario Martínez-Zarzuela:** Conceptualization, Funding acquisition, Investigation, Methodology, Project administration, Software, Supervision, Writing – original draft.

**Funding**

This research has been partially funded by the Department of Employment and Industry of Castilla y León (Spain), under research project ErgoTwyn [INVESTUN/21/VA/0003], and by the Ministry of Science and Innovation (Spain) [PID2021-124515OA-I00].

**Declaration of competing interest**

The authors declare that they have no known competing financial interests or personal relationships that could have appeared to influence the work reported in this paper.

**Acknowledgements**

This study was made possible through a collaboration with the occupational risk prevention department of EuroBelt Valladolid. We would like to thank them and EuroBelt workers for their professionalism during data collection. We would also like to highlight the support they provided, which allowed us to carry out the work in real industrial workplaces.

**Data availability**

The data that has been used is confidential.